\documentclass{revtex4}
\usepackage{graphicx}
\begin{document}
\preprint{P1WG2_diaz_9999}
\title{Search for the Lepton Flavour Violating Higgs decay $H\to \tau\mu$
      at Hadron Colliders}
\author{U. Cotti$^a$}
 \email[ucotti@zeus.umich.mx]{}
\author{L. Diaz-Cruz$^b$}
 \email[ldiaz@sirio.ifuap.buap.mx]{}
\author{C. Pagliarone$^c$}
 \email[pagliarone@fnal.gov]{}
\author{E. Vataga$^c$}
 \email[vataga@fnal.gov]{}
\affiliation{a)IFM-UMSNH, A.P. 2-82, 58041 Morelia, M\'exico}
\affiliation{b) Instituto de F\'{\i}sica, Benem\'erita Universidad Aut\'onoma
de Puebla, Ap. Postal J-48, 72500, Puebla, Pue., M\'exico}
\affiliation{c) INFN, Pisa, Italy}

\begin{abstract}
We study the prospects to detect at hadron colliders the Lepton Flavour
Violating Higgs decay $H\to \tau\mu $, which can reach substantial
branching fractions in several extensions of the SM. Among them, the
generic two higgs doublet model (THDM-III) can be taken as a
representative case where  $B.R.(H\to \tau \mu )$ can reach values of
order $\simeq 10^{-1}-10^{-2}$. Bounds on the LFV factor
$\kappa_{\tau\mu}$ of order $0.8-1.7$ can be derived at 95 \% c.l. at
Tevatron Run-2 with 4 $fb^{-1}$ for $m_H=110-150$ GeV.
\end{abstract}

\maketitle

\section{Introduction.}
The search for the Higgs boson is one of the main goals of Tevatron
RUN-2 and future colliders~\cite{Carena:2000yx}. Although the most
conservative search strategy uses the theoretical expectations coming
from the minimal standard model (SM), it is certainly worthwhile to
look for other signals arising from physics beyond the SM. In this
regard, it has been recognized recently that the Higgs sector of
several well-motivated  models can predict  lepton flavour violating
(LFV) Higgs decays at sizable rates~\cite{Diaz-Cruz:1999xe}, that may
be detectable at future colliders~\cite{Sher:2000uq,Han:2000jz}.
 In fact, violation of Lepton number is predicted
in several extensions of the standard model (SM), and the results
on atmospheric neutrinos~\cite{Fukuda:1998mi}, showing evidence for
neutrino oscillations, indicate that a lepton sector beyond the SM is
required to account for the pattern of neutrino masses and mixing
suggested by the data.

In this paper we report preliminary work started at Snowmass 2001,
regarding the detectability of such LFV Higgs decays at Hadron
colliders. Although such decays were studied
in~\cite{Diaz-Cruz:1999xe}, within the context of several extensions of
the SM, including both the effective lagrangian approach as well as
specific models, we shall express our results following the general
Two-Higgs doublet model (THDM-III), which does serves as a prototype
model where such effects are predicted and it also facilitates the
presentation of bounds on the LFV couplings. The main result of this
work concerns the decays $H\to  \tau \mu$, which was found to reach a
B.R. of order $0.1-.01$, for both the effective lagrangian and THDM-III
cases, which can be searched at Tevatron Run-II.

\section{The LFV Higgs decays in the THDM-III}
The two-Higgs doublet extension of the SM (THDM), (Models I and
II)  solved their problem with large Flavour changing neutral scalar
interactions (FCNSI), by requiring a discrete symmetry that
restricted each fermion to couple at most to one Higgs
doublet~\cite{Glashow:1977nt}. Later on, it was found that FCNSI could be
suppressed at acceptable rates, with relatively light Higgs
bosons, by imposing a more realistic pattern on the Yukawa
matrices, which in principle can be associated with some family
symmetry~\cite{Sher:1998sj}. The phenomenological predictions of
this model (called model III in the
literature~\cite{Atwood:1997vj,DiazCruz:1995er})
have been studied to some extent.

Working within the Higgs mass-eigenstate basis,
 the LFV interactions of the light neutral
Higgs boson H take the form:
\begin{equation}
 {\cal{L}}_{LFV} = \xi_{ij} \cos \alpha \bar{l_i} l_j H + h.c.
\end{equation}
where $\alpha$ denotes the mixing angle of the neutral Higgs sector, and
$\xi_{ij}$ denotes the Yukawa coupling of the second
doublet.
In order to satisfy the low energy data on FCNC, Cheng and
Sher~\cite{Cheng:1987rs,Antaramian:1992ya} proposed the following ansatz :
\begin{equation}
 \xi_{ij}= \lambda_{ij} \frac{(m_i m_j)^{1/2}}{v}
\end{equation}
where $v=246$ GeV and the lepton mass factor gives the order
of magnitude of the interaction. The coefficients
$\lambda_{ij}$ are dimensionless parameters that
can be constrained by comparing the prediction for
relevant processes with present experimental
bounds on FCNC and LFV transitions.
The strongest bound for the parameters $\lambda_{ij}$ are
obtained from muon anomalous magnetic
moment~\cite{Nie:1998dg},
namely: $  \lambda_{\mu \tau}  < 10$, which involves
only one coupling.
 Other interesting bounds are:
 $( \lambda_{e\mu}\lambda_{\mu \tau} )^{1/2} < 5$, which
is obtained from the decay $\mu \to e \gamma$.

One additional implication of these LFV couplings is the possibility to
observe the LFV Higgs decays $H \to l^+_i l^-_j$, whose decay width
will be given by:
\begin{equation}
 \Gamma ( H \to l^+_i l^-_j)=  \frac{\xi^2_{ij} }{8\pi} \cos^2\alpha~m_H
\end{equation}
Since the dominant decay mode of the Higgs boson in the intermediate
mass range is $H \to b \bar b$, and the corresponding coupling will
depend on $m_b$, $\sin^2 \alpha$ and $\lambda_{bb}$, this will
introduce a complicated expression for the Higgs total width. However,
if we assume values for the parameters $\lambda_{\tau\mu}$ and
$\sin\alpha$ of order one, which satisfy present low-energy bounds, and
neglect the corrections to all other higgs decays, we find that $H \to
\mu \tau$ is allowed to have a  B.R. of order $0.1-0.01$, which seems
at the reach of future colliders.
 On the other hand, the resulting upper limit on B.R.$(h^0 \to e \mu)$
is of the order $10^{-5}-10^{-6}$, which does not
seem to be at the reach of future experiments.
 Results for the LFV higgs decays $H \to \tau\mu$ and $H \to \tau e$,
are shown in table 1. Within the minimal SUGRA-MSSM  and the SM
with massive neutrinos, these decays are found to have negligible
rates. Whereas in models with heavy majorana  neutrinos, the LFV
Higgs decays are induced at one-loop level and can reach values of
order $10^{-3}$~\cite{Pilaftsis:1992st}.

\section{Strategy search for $H \to \tau \mu$}
In order to study these LFV Higgs decays at future hadron colliders, we
shall focus on the mode $H \to \tau\mu$, which has the muon in the
final state and it is easier to separate from the backgrounds. Then, to
derive bounds on the LFV parameters we shall consider the THDM-III, but
for this it would be necessary to take into account not only the
dependence of the Higgs decay width on these parameters, but also the
full expression for the branching ratio. In order to handle such
multi-parameter dependence, we shall introduce the parameter
$\kappa_{\mu\tau}$, in such a way that the branching ratio for the
decay $H \to \tau\mu$, is given by:
\begin{equation}
 B.R. (H \to \tau \mu)= \kappa^2_{\tau\mu}
\left(\frac{2 m_\mu }{m_\tau}\right) B.R.(H\to \tau^+\tau^-)
\end{equation}
where the dependence on the lepton masses has been made explicit,
whereas the dependence on  the parameters $\lambda_{ij}$
and $\alpha$ has been absorbed into the couplings
$\kappa_{\tau\mu}$.

In order to study the possibility to detect the LFV higgs decays, one
can use the gluon-fusion mechanism to produce a single Higgs boson;
assuming that the production cross-section is of similar strength to
the SM case, about 1.2 pb for $m_H=100$ GeV, it will allow to produce
12,000 Higgs bosons with an integrated luminosity of 10 $fb^{-1}$.
Thus, for $B.R.(H\to \tau \mu/\tau e) \simeq 10^{-1}-10^{-2}$  Tevatron
can produce $1200-120$ events. Then, to determine the detectability of
the signal,
 we need to study the main backgrounds to the $H \to \tau\mu$ signal, which
are dominated by Drell-Yan tau pair and WW pair production.
 In Ref.~\cite{Han:2000jz} it was proposed to reconstruct the
hadronic and electronic tau decays, assuming the following cuts: i) For
the transverse muon and jet momentum: $p^\mu_T > m_H/5$, $p^{\pm}_T >
10$ GeV, ii) Jet rapidity for Tevatron (LHC): $|\eta| < 2 (2.5)$ iii)
The angle between the missing transverse momentum and the muon
direction: $\phi(\mu \pm)> 160^o$.

\begin{figure}
\includegraphics[scale = .55, angle = -90]{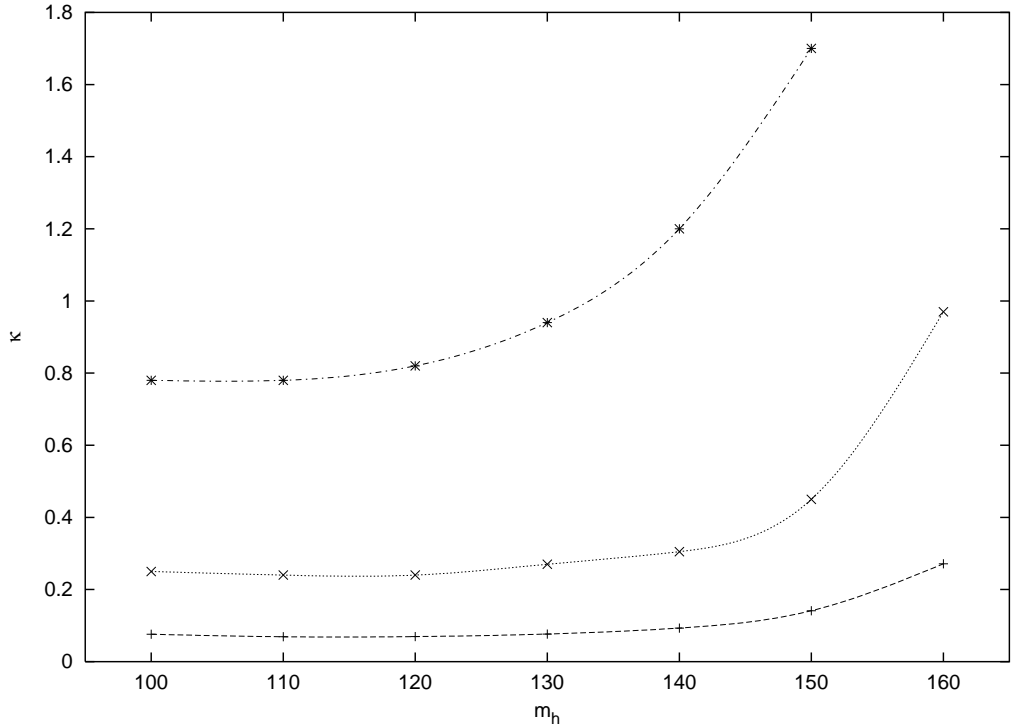}%
\caption{Bounds on the LFV coupling $\kappa_{\tau\mu}$ that can
be obtained at Tevatron Run-2 (dot-dashes), LHC (dots) and VLHC
(dashes).}
\label{kappa}
\end{figure}

The resulting bounds on the LFV higgs couplings $\kappa_{\tau\mu}$ that
can be obtained at Run-2 and LHC at 95\% c.l., are shown in
Fig.~\ref{kappa}; one can see that it will be possible to test values
of $\kappa_{\tau\mu}$ of order  $0.8 - 1.7$ ($0.2-0.4$) at Tevatron
(LHC) with 4 (100) $fb^{-1}$ for $m_H=110-130$ GeV.
 In Fig.~\ref{kappa} we have also included the expected bound on
$\kappa_{\tau\mu}$ at the very large hadron collider (VLHC),
with c.m. energy of 40 TeV and integrated luminosity of
1000 fb$^{-1}$, under the very crude assumption that signal
and background can be scaled from LHC results; in this case
the sensitivity extends up to values of $\kappa_{\tau\mu}=0.1$.

Thus, we found that the LFV Higgs decays $H\to \tau \mu/\tau e$ can
have large branching ratios, of order $0.1-0.01$, in some extensions of
the SM, which can be detected at the coming stages of Tevatron Run-II.
At present we are studying the signal at Tevatron Run-2 using the
realistic detector simulation of CDF, which we expect to present in the
near future~\cite{ucottiinpr}.

\begin{acknowledgments}
Valuable discussions with T. Han, as well as his interest in
helping to establish the contacts, are acknowledged. This work
was also supported by CONACYT, SNI and CIC-UMSNH (M\'exico).
\end{acknowledgments}

\begin{table}
\caption{B.R. of LFV Higgs decays for the THDM-III.
Results are shown for $\sin\alpha=0.1$, and
the numbers in parenthesis correspond to $\sin\alpha=0.9$.}
\label{br}
\begin{tabular}{|l|l|l|}
\hline
$m_H$ GeV & $B.R.(H\to \mu \tau )$  &  $B.R.(H\to e \mu)$ \\
\hline
 100. &   0.7 (0.1)     &  $1.3\times 10^{-5}$ ($2.0\times 10^{-6}$) \\
\hline
 130  &  0.7 (0.1)      &  $1.2\times 10^{-5}$ ($2.1\times 10^{-6}$) \\
\hline
 170. &  0.3 ($1.2\times 10^{-3}$) & $5.5 \times 10^{-6}$
 ($2.3 \times 10^{-8}$) \\
\hline
 200. &  0.1 ($3.5 \times 10^{-4}$) & $2.2 \times 10^{-6}$
($6.4 \times 10^{-9}$) \\
\hline
\end{tabular}
\end{table}

\end{document}